%% file: paper.tex
\documentclass[11pt]{article}
\usepackage{graphicx}
\usepackage{epsfig}

\newcommand{\BABARPubYear}    {00}
\newcommand{\BABARConfNumber} {05}

\newcommand{\SLACPubNumber} {8527}

\input pubboard/babarsym

\newcommand{\kl}{\KL}
\newcommand{\bpsikstar}{\ensuremath{B \to \jpsi \Kstar}}
\newcommand{\thetab}{\ensuremath{\theta_B}}
\newcommand{\cosb}{\ensuremath{|\cos \thetab|}}
\newcommand{\marginkl}{}

\newcommand{\text}[1]{\mbox{\rm #1}}

\newcommand{\BNotJKS}{\Bz$\rightarrow$\jpsi\KS(\pipi)}
\newcommand{\BNotJKSpiz}{\Bz$\rightarrow$\jpsi\KS(\piz\piz)}
\newcommand{\BNotJKz}{\Bz$\rightarrow$\jpsi$K^0$}
\newcommand{\BpJKp}{\Bu$\rightarrow$\jpsi\Kp}
\newcommand{\BNotPsiKs}{\Bu$\rightarrow$\psitwos\Kp}
\newcommand{\BpPsiKp}{\Bz$\rightarrow$\psitwos\KS(\pipi)}
\newcommand{\BNotJKstarz}{\Bz$\rightarrow$\jpsi\Kstarz}

\newcommand{\BpJKstarp}{\Bu$\rightarrow$\jpsi\Kstarp}
\newcommand{\BNotpsitwosKs} {\Bz$\rightarrow$\psitwos\KS}
\newcommand{\BNotpsitwosKz} {\Bz$\rightarrow$\psitwos$K^0$}
\newcommand{\BppsitwosKp} {\Bu$\rightarrow$\psitwos\Kp}

\newcommand{\BpChiKp} {\Bu$\rightarrow$\chicone\Kp}

\newcommand{\az}{A_{0}}
\newcommand{\ap}{A_{\parallel}}
\newcommand{\at}{A_{t}}  

\newcommand{\azd}{|\az|^{2}}
\newcommand{\apd}{|\ap|^{2}}
\newcommand{\atd}{|\at|^{2}}

\def\sb{\sin 2\beta}

\def\chicone{\ensuremath{\chi_{c1}}}

\newcommand{\toEE} {$\rightarrow e^+e^-$}
\newcommand{\toMuMu} {$\rightarrow \mu^+\mu^-$}
\newcommand{\toLL} {$\rightarrow \ell^+\ell^-$ }
\newcommand{\hel} {\ensuremath{|\cos\theta_H |}}
\newcommand{\cosT} {\ensuremath{|\cos\theta_T|}}
\newcommand{\DE}{\mbox{$\Delta E$}}

\long\def\inst#1{\par\nobreak\kern 4pt\nobreak
    {\it #1}\par\vskip 10pt plus 3pt minus 3pt}

\begin{document}
{\pagestyle{empty}

\begin{flushright}
\babar-CONF-\BABARPubYear/\BABARConfNumber \\
SLAC-PUB-\SLACPubNumber
\end{flushright}

\par\vskip 5cm

\begin{center}
\Large \bf Exclusive \boldmath $B$ decays to charmonium final states
\end{center}
\bigskip

\begin{center}
\large The \babar\ Collaboration\\
\mbox{ }\\
July 25, 2000
\end{center}
\bigskip \bigskip

\begin{center}
\large \bf Abstract
\end{center}
We report on exclusive decays of $B$ mesons into final states containing
charmonium using data collected with the 
\babar\ detector at the PEP-II storage rings.
The charmonium states considered here  are \jpsi, \psitwos, and 
$\chicone$. 
Branching fractions for several exclusive final
states, a measurement of the decay amplitudes for
the \bpsikst\ decay, and 
measurements of the \Bz\ and \Bu\ masses are presented. All of the results we
present here are preliminary.

\vfill
\centerline{Submitted to the XXX$^{th}$ International Conference on High Energy
Physics, Osaka, Japan.}
\newpage
}

\input pubboard/authors

\setcounter{footnote}{0}

\section{Introduction}
\label{sec:Introduction}
An understanding of the decays of $B$ mesons to final states including
a charmonium resonance
(\jpsi, \psitwos, $\chicone$) 
is a prerequisite to an analysis
of \CP\ violation in the $B$ system.
In this paper we report 
the measurement of several branching fractions 
of exclusive decays, some of which have been used in our 
measurement of $\sb$~\cite{ref:sin2b}. The channels considered are listed in 
Table~\ref{tab:modes}.
Here and throughout this paper the inclusion of charge conjugate states
is implied.

\begin{table}[tbh]
\caption[tab:modes]{$B$ meson decay modes considered in this paper.}
\begin{center}
\begin{tabular}{|l|l|} \hline
        Channel & Secondary decay mode(s) \\ \hline\hline
\Bz$\rightarrow$\jpsi\KS & \jpsi \toLL; \KS $\rightarrow$ \pipi\ , \piz\piz \\
        \BpJKp   & \jpsi \toLL   \\
        \BNotJKstarz & \jpsi \toLL; \Kstarz $\rightarrow$ \Kp\pim\ , \KS\piz   \\
        \BpJKstarp & \jpsi \toLL; \Kstar $\rightarrow$ \KS\pim\ ,  \Kp\piz  \\
        \BNotpsitwosKs & \psitwos \toLL,
 \jpsi\pipi\ ; \KS $\rightarrow$ \pipi \\
        \BppsitwosKp &  \psitwos \toLL,  \jpsi\pipi   \\
        \BpChiKp  & \chicone  $\rightarrow$ \jpsi $\gamma$ ; \jpsi \toLL \\ \hline
\end{tabular}
\label{tab:modes}
\end{center}
\end{table}

We have used some of these exclusive modes to 
measure the masses of the \Bu\ and
\Bz\ mesons and their mass difference. 
We also present initial results on the yield of \bpsikl\ which
will be used for a future \CP\ analysis.  Finally we describe  an amplitude
analysis of the $B \rightarrow \jpsi\Kstar$ decay.

\section{The \babar\ detector and dataset}
\label{sec:babar}

The \babar\ detector is located at the \pep2\ storage ring, an $e^+e^-$
facility operating at the Stanford Linear Accelerator Center.
\pep2\ collides 9.0\gev\ electrons with 3.1\gev\ 
positrons to give a center of mass energy of 10.58\gev, the
mass of the \FourS\ resonance.

The \babar\ detector is described elsewhere~\cite{ref:babar}; here we give 
only a brief overview.
Surrounding the interaction point is a 5-layer double-sided
silicon vertex tracker (SVT) 
which gives precision spatial information for
all charged particles, and is the primary detection device for low momentum
charged particles. Outside the SVT, a 40-layer drift chamber (DCH) 
provides measurements of charged particle momenta. The
\dedx\ information 
from the DCH and SVT is used for particle identification. 
Beyond the outer radius of the DCH is a detector of internally reflected 
Cherenkov radiation (DIRC) which is used primarily for charged hadron
identification. The  detector consists of quartz bars in
which Cherenkov light is produced as relativistic charged particles traverse
the material. The light is internally reflected, and the Cherenkov rings
are measured with an array of photo-multiplier tubes mounted on the rear of
the detector. A CsI(Tl) crystal electro-magnetic calorimeter (EMC) is used to 
detect photons and neutral hadrons, 
as well as to identify electrons.
The EMC is surrounded by a super-conducting 
solenoid which produces a 1.5 T magnetic field. The Instrumented Flux Return
(IFR) consists of multiple layers of resistive plate chambers 
interleaved with the flux return iron. It  is
used in the identification of muons and neutral hadrons.

The data used in these analyses 
correspond to a integrated 
luminosity of 7.7\invfb\ taken 
on the $\FourS$ and 1.2\invfb\ taken 0.04\gev\
below the peak. The data set contains  $8.8\times 10^6$ \BB\ events. 
For the analysis of the $B$ meson masses we use a restricted set of data
corresponding to 4.6\invfb.

\section{Particle reconstruction}
\label{sec:Analysis}

Inclusive charmonium reconstruction is described in detail
in another contribution to this conference~\cite{ref:incharm}.

We here reconstruct \jpsi\ candidates by combining pairs of oppositely charged
tracks within the angular range $0.41 < \theta < 2.41 (2.53)$ for electron (muon) candidates, where $\theta$ is the polar angle
to the beam axis.
The invariant mass of the candidate must lie in the range $2.95 < {\rm m}_{\jpsi} < 3.14$\gevcc\ and 
$3.06 < {\rm m}_{\jpsi} < 3.14$\gevcc\ for decays to $e^+e^-$ and $\mu^+\mu^-$, respectively.
When the \jpsi\ decays to electrons, we demand that at least one of
the tracks pass stringent particle identification requirements based
on the ratio of the energy deposited in the EMC to the track momentum
($E/p$), and on the ionization loss of the track in the drift chamber (\dedx).
For $\mu^+\mu^-$ candidates, 
one track is required to pass a loose muon selection on the basis of 
the number of hit layers in the IFR and the other track is 
required to have an associated energy in the EMC which is
consistent with a minimum ionizing particle. In the case of the decay
to electrons, we apply a procedure to add photons 
which are close to the electron 
tracks and thereby reduce the impact of bremsstrahlung on the reconstruction
efficiency.

We select $\psi(2S)\rightarrow \ell^+\ell^-$ candidates in a similar way. 
For the decay to  $\mu^+\mu^-$ the invariant mass of the candidate is required 
to be within 0.05\gevcc\ of the nominal mass.
In the case of $\psi(2S)$ decays to $e^+e^-$ the lower limit is relaxed to
0.25\gevcc\ below the nominal mass value.
For the decay \psitwos\ to $J/\psi\pi^+\pi^-$,  \jpsi\ candidates are
combined with pairs of oppositely charged tracks which originate
from a common vertex, and the mass difference between the resulting
\psitwos\ candidate and the \jpsi\ is required to be within 0.05\gevcc\ of the
nominal mass difference. 
For the decay \jpsi\toEE, the mass difference is relaxed to 
$-0.25 < {\rm m}_{\psitwos} - {\rm m}_{\jpsi} < 0.05$\gevcc.

Candidate \chicone\ mesons are reconstructed via their decay to \jpsi$\gamma$. 
The $\gamma$ candidates are selected by requiring a neutral cluster in the EMC
that has a distribution of crystal energies consistent with
a $\gamma$ shower.
 The mass difference between the
reconstructed \chicone\ and the \jpsi\ is required to satisfy $0.35<\Delta M
< 0.45$\gevcc, and the momentum of the \chicone\ in the \FourS\ rest frame must
lie in the range $1.15 < p^* < 1.70$\gevc.

$\KS\rightarrow\pipi$ candidates are 
formed from pairs of oppositely charged tracks 
which have an invariant mass between 0.489 and 0.507\gevcc.
$\KS\rightarrow\piz\piz$ candidates are required to have a mass between
0.470 and 0.525\gevcc\ and an energy greater than 0.8\gev. 
A \piz\ decay to two photons is observed in the EMC either as a single neutral cluster with substructure
or as two distinct $\gamma$ clusters.
The most probable decay point of the \KS\ is determined after refitting the
two \piz\ mesons at several points along the path defined by their summed momentum 
vector and the \jpsi\ vertex.  


We reconstruct \Kstarz\ decays to \Kp\pim\ and \KS\piz, and  \Kstarp\ decays to
\KS\pip\ and \Kp\piz. In all cases the candidate $K^{*}$ is required to have an
invariant mass within 0.075\gevcc\ of the nominal value.

A \KL\ candidate is reconstructed using neutral clusters
observed in the EMC or the IFR.  For an EMC candidate
we require the deposited energy to be between 0.2 and 2.0\gev\
and the cluster center-of-gravity to be well contained within the fiducial volume of the
calorimeter ($\cos\theta<0.935$). We reject candidates that are likely to be produced by
photons by means of energy-dependent criteria based on the spatial
distribution of the deposited energy in the cluster.  A neutral cluster
that can be combined with other neutral clusters to form a \piz\ or
clusters with sub-structure consistent with a \piz\ is also rejected.  
A \KL\ candidate observed in the IFR is required to have the cluster center
within the fiducial volume $(-0.75 < \cos\theta < 0.93)$ and to have
a signal in at least two detector layers. We also apply additional
isolation criteria to remove candidates that may have been split from
clusters produced by charged particles.

\section{Exclusive {\boldmath $B$} reconstruction}
For the decays \BNotJKS\ and \BpJKp, we require \hel, the absolute value of
the cosine of the helicity angle of the \jpsi, to be less than
0.8 (0.9) for \jpsi\ decays to $e^+e^-$ ($\mu^+\mu^-$).  In addition we 
require that the \KS\ be consistent with
having originated from the \jpsi\ vertex.

For the decay \BNotJKSpiz\ we require $\hel < 0.75$ (0.8) for \jpsi\
decays to $e^+e^-$ ($\mu^+\mu^-$). To further reduce background in the 
$J/\psi\rightarrow e^+e^-$ channel, we require both tracks to satisfy 
electron identification criteria, one stringent and one loose.

In reconstructing the decays \BNotJKstarz\ and \BpJKstarp\ we require
that a candidate charged kaon satisfy particle identification criteria based on ionization 
loss 
in the DCH and SVT and on the Cherenkov angle measured in the DIRC. 
The candidate \piz\ mass is required to lie in the range
0.115 to 0.150\gevcc. We require $\cosT<0.9$, where $\theta_T$ is the 
angle between the thrust direction of the reconstructed B and that of
the rest of the event in the \FourS\ rest 
frame.  For the decay
$\Kstarz\rightarrow\Kp\pim$, the vertices of the \Kstarz\ and \jpsi\ must be
consistent with a single production point.

For the decays \BNotPsiKs\ and \BpPsiKp\ we require
$\cosT <0.9$ for $\psi(2S) \rightarrow  J/\psi \pi^+\pi^-$ decays and 
$\hel < 0.8$ for $\psi(2S) \rightarrow \ell^+\ell^-$ decays.  We also
require the \KS\ flight length to be greater than $2.5$\mm\ and the \Kp\
to satisfy loose  kaon identification criteria.

We reconstruct the decay \BpChiKp\ with the requirement $\cosT <0.9$ and
demand that the \Kp\ pass loose kaon identification criterion to further reduce
the background in this channel. 

For \marginkl \bpsikl\ decays we require \hel\ and \cosb\ to be less
than 0.9, where \thetab\ is the angle of the $B$ candidate direction with
respect to the beam axis in the rest-frame of the \FourS.  We also
require the sum of these quantities to be less than 1.3.

To isolate the signal for each mode we use the variables \DE, the
difference between the reconstructed and expected $B$ meson energy 
measured in the center-of-mass frame, and \mes, the beam-energy substituted
mass. These variables are defined as:
\begin{eqnarray}
\mes &=& \sqrt{E^{*2}_b - \mbox{\boldmath $p$}_B^{*2}},  \\
\DE &=& E_B^* - E_b^*, 
\end{eqnarray}
where $E_b^*$ is the beam energy in the center-of-mass, i.e., half the
center-of-mass energy, and $E_B^*$ and
$\mbox{\boldmath $p$}_B^*$ are the energy and momentum of the
reconstructed $B$ meson in the center-of-mass.
For \marginkl the \bpsikl\ selection
the momentum of the \kl\ candidate is obtained by constraining the invariant mass of the
\kl\ and \jpsi\ combination to the mass of the $B$ meson.  Therefore, in this
mode only \DE\ can be used to separate the signal from background.
We exclude any event that passes the
other exclusive $B$ selections,  has \mes\ greater than
5.2\gevcc\ and  $|\DE|$ less than 0.1\gev.

Only one exclusive candidate per event is accepted.  If there are
multiple candidates, we select the one with the smallest 
value of $|\DE|$.
Exceptionally, \marginkl
in the \bpsikl\ selection
we choose the candidate with the largest \kl\ energy as
measured by the EMC.  If none of the candidate  \kl\ mesons have EMC information, we choose 
the candidate that has 
the largest number of layers with hits in the IFR.

\section{Results}
\label{sec:Physics}
\subsection{Branching fraction measurements}
\label{sec:excl}
When deriving branching fractions 
we have used the secondary branching fractions and their
associated errors published by the Particle Data Group~\cite{ref:PDG}.

We determine the number of \BB\ events from the difference in the
multi-hadron rate  on and off the \FourS\ resonance, normalized
to the respective luminosity. This leads to a systematic error of 3.6\%
on all measured branching fractions.
We have assumed the branching fraction of the \FourS\   
to \BB\ is 100\%,
with an equal admixture of charged and neutral $B$ final states.

The efficiencies for each mode have been obtained from Monte Carlo simulations
complemented with measurements of tracking and particle identification
efficiencies extracted from the data.
From particle identification control samples we assign a systematic error of 2\% (3\%) per electron (muon).
We attribute a 5\% systematic error to the  \piz\ reconstruction 
efficiency and resolution.
The track finding efficiency has an uncertainty of 2.5\% per track.
Uncertainties in the modeling of the track resolution lead to an additional
1--2\% error depending on the details of the primary and secondary decays.

For each mode the shape of the beam-energy substituted mass distribution 
is  parameterized with the sum of a Gaussian and the ARGUS function~\cite{ref:ARGUS}.
We assign a systematic error due to our uncertainty of the shape of the
background of between 1\% and 9\% depending on the mode.
For the \bpsikstar\ channels, a likelihood fit is
performed for all the decay modes simultaneously, taking into account
the cross-feed between decays.

\begin{figure}
\begin{center}
\mbox{\epsfig{figure=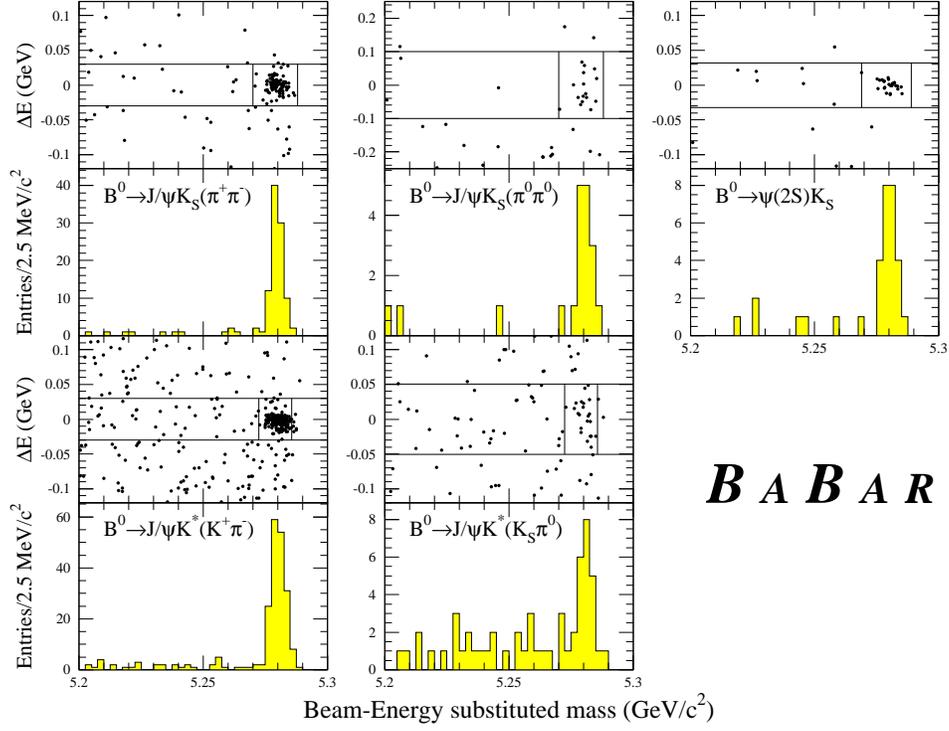, width = 5.0in}}
\mbox{\epsfig{figure=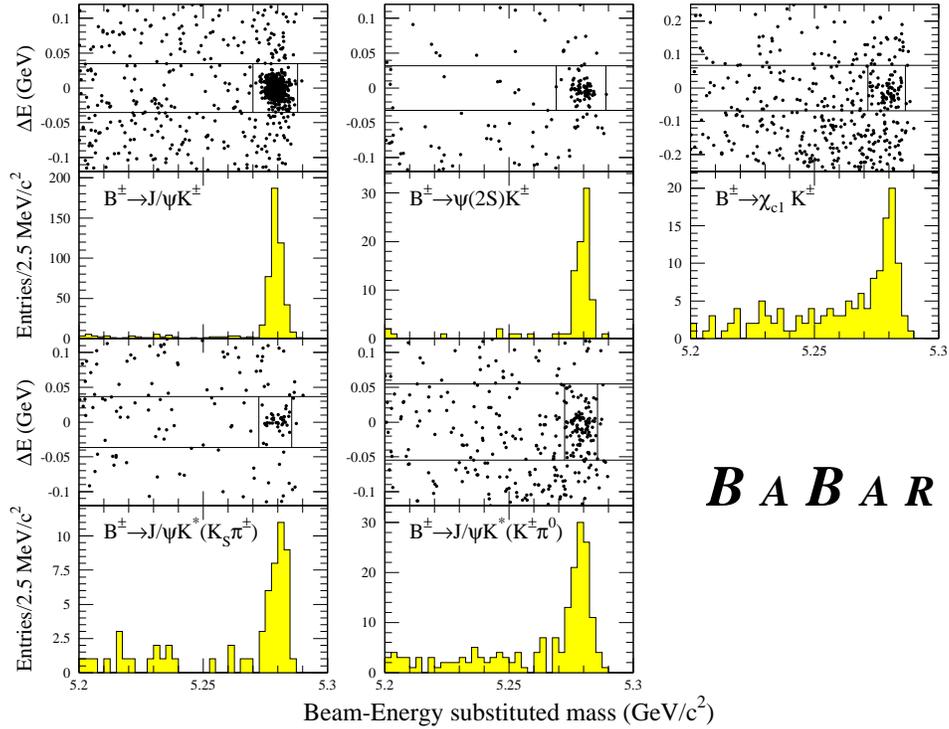, width = 5.0in}}
\caption{\label{fig:deandmb} 
Distributions of candidate events in \mes\ and 
\DE. The upper plot shows the \Bz\ modes and the lower plot the \Bu\
modes. All results are preliminary.}
\end{center}
\end{figure}

\begin{figure}
\begin{center}
\mbox{\epsfig{figure=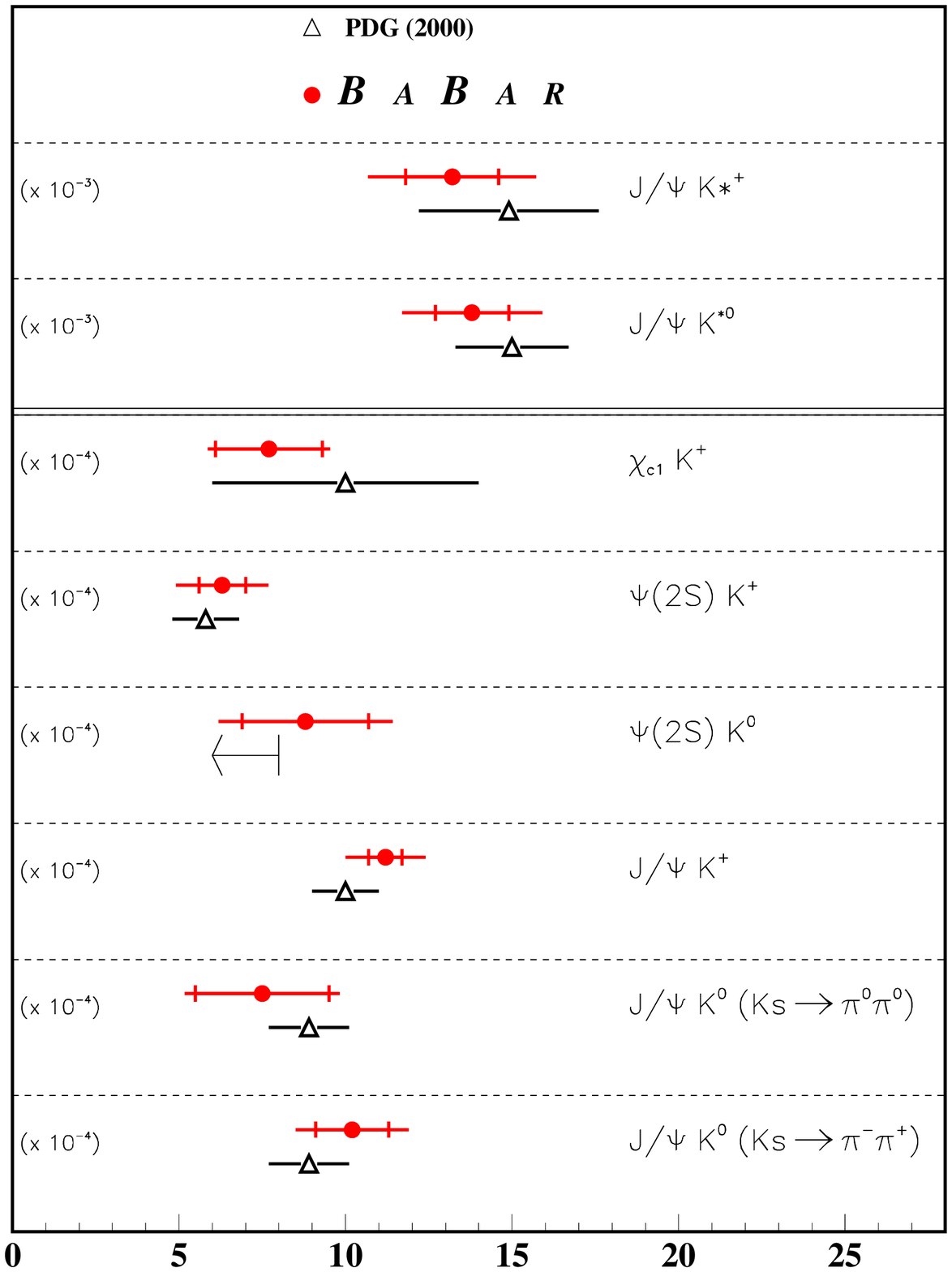, height = 7.5in}}
\caption{\label{fig:brsum} 
Summary of branching fraction measurements for charmonium + $K$ channels 
and comparsons with the PDG 2000 values. All results are preliminary.}
\end{center}
\end{figure}

Figure~\ref{fig:deandmb} shows the \mes\ and \DE\ 
distributions of the candidates.
In Table~\ref{tab:yield} we present the yields 
and measured branching fractions for the individual exclusive
modes.
Figure~\ref{fig:brsum} shows the measured branching fractions compared
to the values compiled by the Particle Data Group~\cite{ref:PDG}.

\begin{table}[hb]
\caption{\label{tab:yield}The yields and measured branching fractions
for exclusive decays of $B$ mesons involving charmonium. The yield only
includes the statistical error. For the branching fractions, 
the first error is statistical and the second systematic. All results are preliminary.}\medskip
\begin{center}
\begin{tabular}{|llcc|}
\hline
Channel &   & Yield   & Branching fraction/$10^{-4}$\\
\hline\hline
\BNotJKz    &  \KS\ $\rightarrow \pi^+\pi^-$   & 93 $\pm$ 10 & 
$10.2 \pm 1.1 \pm 1.3 $  \ \\
             &  \KS\ $\rightarrow \pi^0\pi^0$ & 14 $\pm$ 4 & 
$7.5 \pm 2.0 \pm 1.2  $ \ \\
\BpJKp &            & 445 $\pm$ 21 & 
$11.2 \pm 0.5 \pm 1.1  $ \ \\
\BNotJKstarz &      & 188  $\pm$ 14 & 
$13.8 \pm 1.1 \pm 1.8  $ \ \\
\BpJKstarp &  & 126 $\pm$ 12  & 
$13.2 \pm 1.4 \pm 2.1  $ \ \\
\BNotpsitwosKz & & 23 $\pm$ 5 & 
$8.8 \pm 1.9 \pm 1.8  $ \ \\
\BppsitwosKp & & 73  $\pm$ 8 & 
$6.3 \pm 0.7 \pm 1.2  $ \ \\
\BpChiKp & & 44 $\pm$ 9 & 
$7.7 \pm 1.6 \pm 0.9  $ \ \\
\hline
\end{tabular}
\end{center}
\end{table}

\subsection{Observation of a signal for the decay \boldmath $\bpsikl$}
\begin{figure}
\begin{center}
\mbox{\epsfig{figure=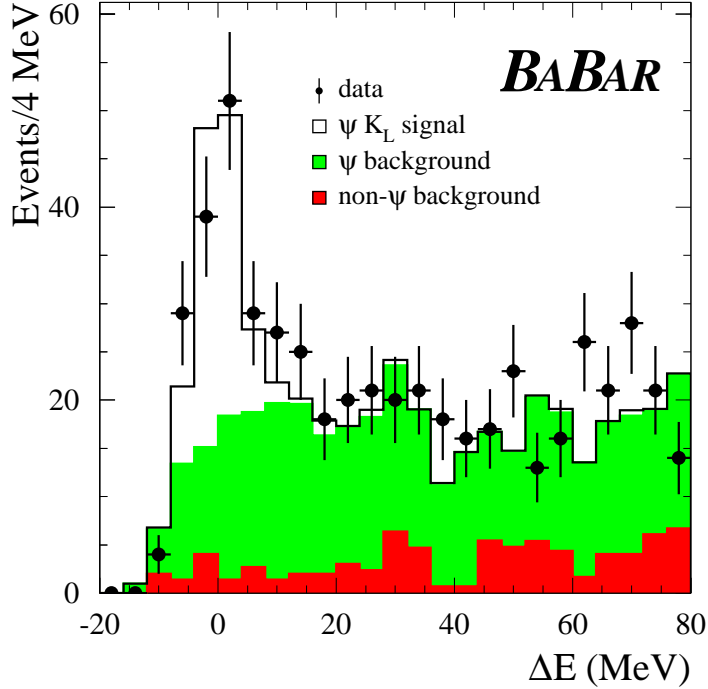,height=4.0 in}}
\caption{\label{fig:psikl} Comparison of data with Monte Carlo
simulation for the \bpsikl\ selection. The results presented here are preliminary.}
\end{center}
\end{figure}

Figure~\ref{fig:psikl} shows the \marginkl \DE\ distributions for the
\bpsikl\ candidates in data and Monte Carlo events. We determine the yield by
counting events with $\DE < 0.01$\gev\ and subtracting the background
contributions. There are two categories of background to this mode.
The first arises from other $B\rightarrow\jpsi X$ decays and is
estimated from simulations normalized to the measured \jpsi\ yield. We
estimate the systematic error on the yield from this background determination 
by varying the \kl\ reconstruction efficiency, the branching fractions
of the major background modes, and the helicity amplitudes used in the
simulation of \bpsikstar\ decays. The second background arises from
non-\jpsi\ modes and is measured from a sideband above the \jpsi\ peak
in the data.  The systematic error from this contribution is
determined by varying the shape of the background.
We measure the yield of \bpsikl\ events to be $82 \pm 14$ (stat) $\pm 9$ (syst).  
This is in good agreement with the expected number of 93
signal events predicted from Monte Carlo simulation.

\subsection{Measurements of the \boldmath $\Bz$ and $\Bu$ masses}

Measurements of the $\Bz$ and $\Bu$ invariant masses have been
performed using the decay modes 
\BNotJKS, \BNotJKstarz(\Kp\pim) and \BpJKp. These modes are chosen
because they have small backgrounds and the 
masses of the secondary decay products are well known.
The event selections are as described in section~\ref{sec:excl}, with the
additional requirements  $\mes > 5.27$\gevcc\ and 
$|\DE| < 36$\mev.

The $B$ candidate invariant mass is derived by fitting the decay products to
a common vertex, with the 
masses of the \jpsi\ and \KS\ constrained to their nominal values.
Uncertainties in the magnetic field and in the internal and relative
alignment of the tracking devices can introduce a bias in the momentum
measurement. The size of this effect is quantified by comparing the
reconstructed mass of \jpsi\toMuMu\ candidates, determined by fitting to the
invariant mass distribution, to the nominal mass value. Any observed shift 
is subsequently applied to the track momenta in simulated data to determine a
correction to the measured $B$ mass. The systematic error attributed to this
correction is derived from statistical uncertainty on the parameters from the
fit to the \jpsi\ invariant mass. The \BNotJKS\ sample requires special 
consideration as
the decay products of the \KS\ do not come from the interaction point 
and are sensitive
to details of the track parameterization. A fit to the observed \pipi\ invariant mass is
performed and a correction derived in the same way as described above.
The correction applied to the \BNotJKS\ sample is taken 
to be the mean of the correction factors determined from the fits to the
\jpsi\ and \KS\ 
distributions, taking the semi-dispersion as the error. The resulting
systematic uncertainty is $\pm 0.62$, $^{+0.59}_{-0.62}$ and
$^{+0.42}_{-0.44}$\mevcc\ for the 
\BNotJKS, \BNotJKstarz(\Kp\pim) and \BpJKp\ modes, respectively.

\begin{figure}[tbh]
\begin{center}
\mbox{\epsfig{figure=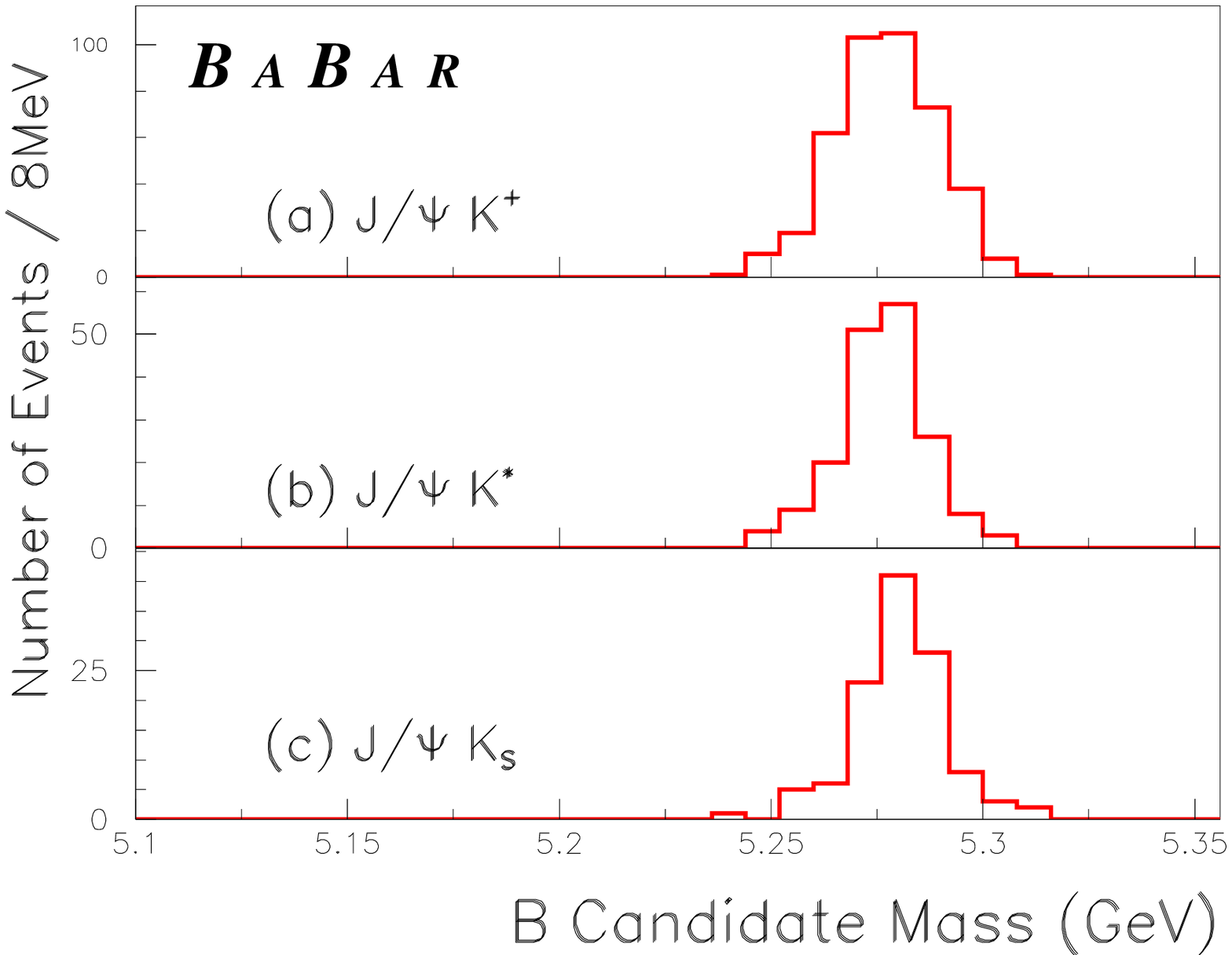, bbllx=5,bblly=0,bburx=517,bbury=407,clip=, height = 3.0in}}
\caption{\label{fig:mass} 
The reconstructed $B$ mass distribution for the (a) 
\BpJKp , (b) \BNotJKstarz(\Kp\pim) and (c) \BNotJKS\ samples. The results
presented here are preliminary.}
\end{center}
\end{figure}

An additional uncertainty comes from background contamination in the 
event samples, which are determined from a fit to the sideband events, and is
found to be between 2\% and 4\%. The measurement of the mass 
is performed by fitting a single Gaussian
and a flat background to the $B$ invariant mass distribution. Examples of the
mass distributions are shown in Fig.~\ref{fig:mass}. 
The distortion of the mass measurement due to the presence of background has
been estimated by removing the $N$ events with
smallest mass and the $N$ events with highest mass, where $N$ is the number of
background events in the sample. The corresponding
systematic uncertainty is  $^{+0.73}_{-0.60}$, $^{+0.62}_{-0.61}$ 
and $^{+0.30}_{-0.28}$\mevcc\ for the 
\BNotJKS, \BNotJKstarz(\Kp\pim) and \BpJKp\ modes, respectively.

The $B$ masses have been measured to be:
\begin{eqnarray*}
m(\Bz) &=& 5279.0 \pm 0.8\ ^{+0.8}_{-0.8}\ \mevcc, \\
m(\Bu) &=& 5278.8 \pm 0.6\ ^{+0.4}_{-0.4}\ \mevcc,
\end{eqnarray*}
where the first error is the quadratic sum of the statistical and
uncorrelated systematic errors and the second error is the correlated
systematic error.

The mass difference between the \Bz\  and  \Bu\ mesons
is evaluated by fitting the \mes\ distributions
of the three above-mentioned channels with the ARGUS function to describe the background
and a Gaussian to describe the signal.
The use of \mes\ has the advantage that it reduces the sensitivity to the
measured momentum scale and the uncertainties in the energy scale of the beam
particles cancel in the mass difference.
The same systematic study has been performed as in the
invariant mass measurement described above, and was found to contribute 0.01\mevcc\ to the 
systematic error on the mass difference measurement.
Simulations indicate that the effect of the uncertainties in the beam parameters 
on the mass difference measurement is only 0.001\mevcc.

We also consider how the uncertainty in the shape of the background under the
signal affects the mass difference measurement.
We estimate the uncertainty
by fitting the shape of the distribution in the \DE\ sidebands and
using these parameters when fitting to the signal. 
The effect on the mass difference between fixing the background shape 
or not is found to be 0.04\mevcc.

We measure the mass difference to be:
\begin{eqnarray*}
m(\Bz)-m(\Bu) = 0.28\pm 0.21 \pm 0.04\ \mevcc 
\end{eqnarray*}
where the first error is the quadratic sum of the statistical and
uncorrelated systematic errors and the second error is the correlated
systematic error.


\subsection{Angular analysis of \boldmath $B\rightarrow \jpsi\Kstar$}
The $B \rightarrow \jpsi\Kstar$ decay proceeds through three
amplitudes, corresponding to the three different helicity configurations of
the decay products \cite{ref:physbook}.
The transversity formalism involves linear combinations of these amplitudes,
denoted by $\az$, $\at$ and $\ap$.
Both $\az$ and $\ap$ are \CP\ even while $\at$ is \CP\ odd.
The size of the \CP\ odd contribution in the decay must be known before a value
of $\sb$ can be extracted from this decay channel.

The event selection is similar to that used for the branching fraction measurement in
section~\ref{sec:excl}.
In this analysis we have considered only those channels which have a
final state composed solely of charged particles.
$B$ candidates were required to have a reconstructed mass within $\pm 0.01$\gevcc\ of the nominal value
and $|\DE| < 0.075$\gev.

The background is determined from the sideband region
$\mes <5.25$\gevcc. 
The amplitudes are determined from a fit using the unbinned extended likelihood 
method~\cite{barlow}, that takes into account the normalization 
condition $\azd+\apd+\atd=1$, the finite detector acceptance and the
background contributions (assumed to have a flat angular distribution).

\begin{figure}[bt] 
\begin{center}
 \mbox{\epsfig{figure=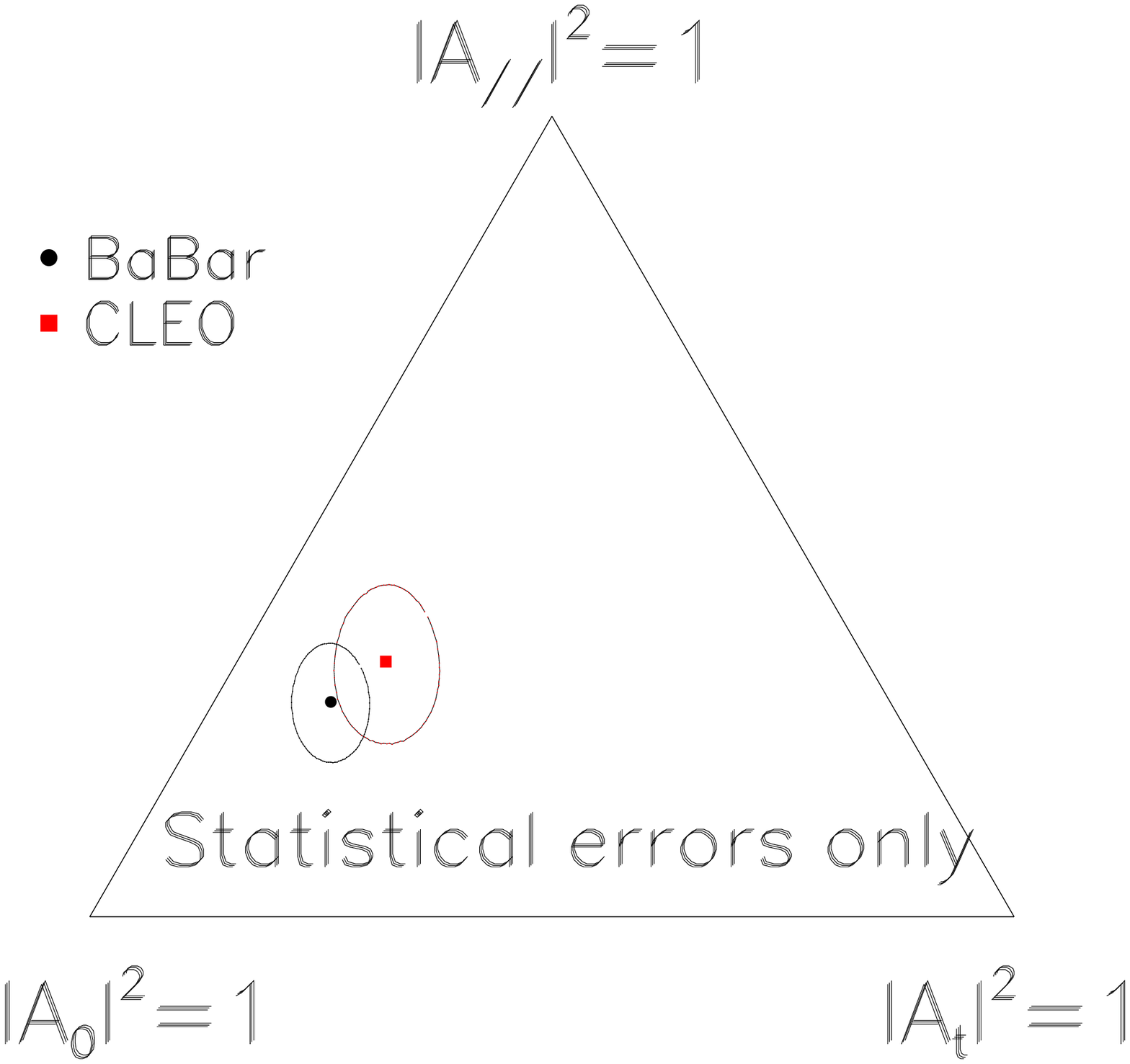,width=0.45\linewidth} }
 \mbox{\epsfig{figure=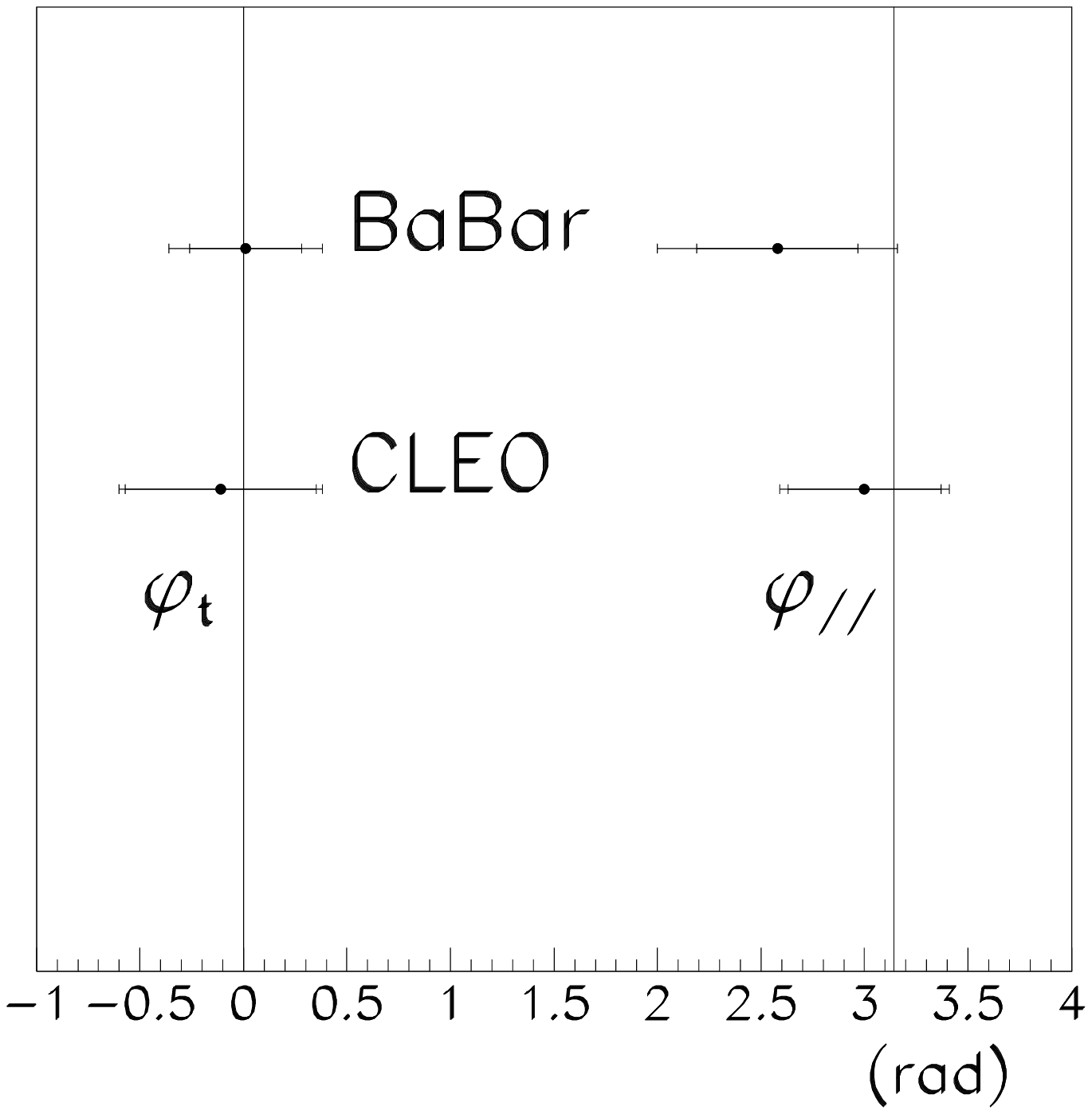,width=0.45\linewidth} }
\caption{Transversity angle fits. 
The modulii of the amplitudes are shown as 68\% contours in the top 
plot while the relative phase of the
amplitudes are displayed in the bottom plot. Also shown are the results
from CLEO~\cite{CLEO97}.
\label{23}}
\end{center}
\end{figure}

The 68\% contours of the fit are presented in the $\azd+\apd+\atd=1$ plane of
the $\azd$, $\apd$, $\atd$ space in Fig.~\ref{23}. The fraction of the
amplitude that is \CP\ odd is determined to be $\atd = 0.13 \pm 0.06 \pm 0.02$, 
while the longitudinal polarization 
($\Gamma_{L} / \Gamma$) is found to be $\azd = 0.60 \pm 0.06 \pm 0.04$.

Systematic errors arising from our knowledge of the background, the
acceptance corrections, the cross-feed among
$B\rightarrow \jpsi\Kstar$ modes, and  the contribution due to heavier
\Kstar\ mesons 
have been considered. 
The effect on the
transversity amplitudes and phases are summarized in Table~\ref{tab:ampsyst}. 

\begin{table}[!htb]
\caption{Systematic uncertainties in the measurement of transversity 
amplitudes.}
\begin{center}
\label{tab:ampsyst}
\begin{tabular}{|l|ccccc|} \hline
Source & $\azd$ & $\atd$ & $\apd$ &
$\varphi_{\parallel}$ (rad) & $\varphi_{t}$ (rad) \\ \hline\hline
Monte Carlo statistics & $0.014$ & $0.014$ & $0.016$ & $ 0.12$ & $ 0.08$ \\ 
Backgrounds & $0.011$ & $0.009$ & $0.001$ & $ 0.01$ & $ 0.03$ \\
Angular acceptance  & 0.020 & 0.011 & 0.020 & 0.13 & 0.05 \\
Cross-feed background & $0.025$ & $0.006$ & $0.030$ & $ 0.02$ & $0.03$ \\
Heavy $\Kstar$ & $0.011$ & $0.004$ & $0.007$ & $0.07$ & $0.01$ \\ \hline
Total & 0.038 & 0.021 & 0.040 & 0.19 & 0.10 \\ \hline
\end{tabular}
\end{center}
\end{table}

The results obtained are presented in Table~\ref{tab:ampres}.

\begin{table}[!htb]
\caption{Measured transversity amplitudes.  The third amplitude is determined
from the normalization condition while the phase $\phi_{0}$ is, by convention,
set to zero.}\medskip
\begin{center}
\label{tab:ampres}
\begin{tabular}{|c|c|c|c|} \hline
$\azd$ & $\atd$ & $\varphi_{\parallel}$ (rad) & $\varphi_{t}$ (rad) \\
\hline\hline
$0.60\pm 0.06 \pm 0.04$ & $0.13\pm 0.06 \pm 0.02$ & $ 2.58\pm 0.39 \pm 0.20$ & $ 0.01\pm 0.27 \pm 0.10$  \\
\hline
\end{tabular}
\end{center}
\end{table}

\section{Summary}
\label{sec:Summary}

We have presented preliminary
measurements of branching fractions of $B$ mesons to several
two body charmonium final states. The results are in good agreement with
previous measurements. 
A signal for the decay \bpsikl\ is observed, with a 
yield compatible with our expectation.

The \Bz\ and \Bu\ masses and their mass difference have been measured, with
results that are in good agreement with the world average values~\cite{ref:PDG}.

Finally we have presented an analysis of the transversity amplitudes in the
decay $B\rightarrow \jpsi\Kstar$ that confirm that this final state is
dominantly \CP\ even and the \CP\ asymmetry measurements in this channel will 
have a small dilution.

\section{Acknowledgments}
\label{sec:Acknowledgments}

\input pubboard/acknowledgements

\end{document}

%% file: pubboard/babarsym.tex
\usepackage{relsize}
\def\babar{\mbox{\slshape B\kern-0.1em{\smaller A}\kern-0.1em
    B\kern-0.1em{\smaller A\kern-0.2em R}}}

\def\piz   {\ensuremath{\pi^0}}

\def\pip   {\ensuremath{\pi^+}}
\def\pim   {\ensuremath{\pi^-}}
\def\pipi  {\ensuremath{\pi^+\pi^-}}

\def\Kbar  {\kern 0.2em\overline{\kern -0.2em K}{}}

\def\Kp    {\ensuremath{K^+}}

\def\KS    {\ensuremath{K^0_{\scriptscriptstyle S}}} 
\def\KL    {\ensuremath{K^0_{\scriptscriptstyle L}}} 
\def\Kstarz  {\ensuremath{K^{*0}}}

\def\Kstar   {\ensuremath{K^*}}

\def\Kstarp   {\ensuremath{K^{*+}}}

\def\Kzb   {\ensuremath{\Kbar^0}}
\def\KzKzb {\ensuremath{K^0 \kern -0.16em \Kzb}}

\def\Dbar  {\kern 0.2em\overline{\kern -0.2em D}{}}

\def\Dzb   {\ensuremath{\Dbar^0}}
\def\DzDzb {\ensuremath{D^0 {\kern -0.16em \Dzb}}}

\def\Bz    {\ensuremath{B^0}}

\def\Bbar  {\kern 0.18em\overline{\kern -0.18em B}{}}

\def\Bzb   {\ensuremath{\Bbar^0}}
\def\Bu    {\ensuremath{B^+}}

\def\BB    {\ensuremath{B\Bbar}} 
\def\BzBzb {\ensuremath{B^0 {\kern -0.16em \Bzb}}}

\def\jpsi  {\ensuremath{{J\mskip -3mu/\mskip -2mu\psi\mskip 2mu}}} 
\def\psitwos {\ensuremath{\psi{(2S)}}}
\mathchardef\Upsilon="7107
\def\Y#1S{\ensuremath{\Upsilon{(#1S)}}}

\def\FourS {\Y4S}

\mathchardef\Deltares="7101
\mathchardef\Xi="7104
\mathchardef\Lambda="7103
\mathchardef\Sigma="7106
\mathchardef\Omega="710A
\def\Deltabar   {\kern 0.25em\overline{\kern -0.25em \Deltares}{}}
\def\Lbar {\kern 0.2em\overline{\kern -0.2em\Lambda\kern 0.05em}\kern-0.05em{}}
\def\Sigbar{\kern 0.2em\overline{\kern -0.2em \Sigma}{}}
\def\Xibar{\kern 0.2em\overline{\kern -0.2em \Xi}{}}
\def\Obar{\kern 0.2em\overline{\kern -0.2em \Omega}{}}
\def\Nbar{\kern 0.2em\overline{\kern -0.2em N}{}}
\def\Xbar{\kern 0.2em\overline{\kern -0.2em X}{}}

\def\bpsikst{\ensuremath{B^0 \to \jpsi \Kstar}}
\def\bpsikl{\ensuremath{B^0 \to \jpsi \KL}}

\def\mes        {\mbox{$m_{\rm ES}$}}

\def\ev   {\ensuremath{\rm \,e\kern -0.08em V}}
\def\kev  {\ensuremath{\rm \,ke\kern -0.08em V}} 
\def\mev  {\ensuremath{\rm \,Me\kern -0.08em V}} 
\def\gev  {\ensuremath{\rm \,Ge\kern -0.08em V}} 
\def\gevc {\ensuremath{{\rm \,Ge\kern -0.08em V\!/}c}} 
\def\tev  {\ensuremath{\rm \,Te\kern -0.08em V}}
\def\mevc {\ensuremath{{\rm \,Me\kern -0.08em V\!/}c}} 
\def\gevcc{\ensuremath{{\rm \,Ge\kern -0.08em V\!/}c^2}} 
\def\mevcc{\ensuremath{{\rm \,Me\kern -0.08em V\!/}c^2}}

\def\mm   {\ensuremath{\rm \,mm}}

\def\invfb   {\ensuremath{\mbox{\,fb}^{-1}}}
\def\mus  {\ensuremath{\rm \,\mus}}

\def\mus        {\ensuremath{\,\mu{\rm s}}}    
\def\gsim{{~\raise.15em\hbox{$>$}\kern-.85em
          \lower.35em\hbox{$\sim$}~}}
\def\lsim{{~\raise.15em\hbox{$<$}\kern-.85em
          \lower.35em\hbox{$\sim$}~}}

\def\CP                 {\ensuremath{C\!P}}

\def\to                 {\ensuremath{\rightarrow}}

\def\pep2{PEP-II}

\newcommand{\dedx}{\ensuremath{\mathrm{d}\hspace{-0.1em}E/\mathrm{d}x}}

\def\sb{${\sin\! 2 \beta   }$}

\newcommand{\eqref}[1]{Eq.~(\ref{eq:#1})}

\newcommand{\epjc}      [1]  {{Eur.\ Phys.\ Jour.\ C~{\bf #1}}}

\newcommand{\nim}       [1]  {{Nucl.\ Instr.\ and Methods~{\bf #1}}}

\newcommand{\pl}        [1]  {{Phys.\ Lett.\ {\bf #1}}}      

\newcommand{\prl}       [1]  {{Phys.\ Rev.\ Lett.\ {\bf #1}}} 

\def\jetset74   {\mbox{\tt Jetset \hspace{-0.5em}7.\hspace{-0.2em}4}}

%% file: pubboard/authors.tex
\begin{center}
\small

The \babar\ Collaboration
\bigskip

B.~Aubert,
A.~Boucham,
D.~Boutigny,
I.~De Bonis,
J.~Favier,
J.-M.~Gaillard,
F.~Galeazzi,
A.~Jeremie,
Y.~Karyotakis,
J.~P.~Lees,
P.~Robbe,
V.~Tisserand,
K.~Zachariadou
\inst{Lab de Phys.\ des Particules, F-74941 Annecy-le-Vieux, CEDEX, France}
% ...

\end{center}\newpage

%% file: pubboard/acknowledgements.tex
We are grateful for the contributions of our \pep2\ colleagues in
achieving the excellent luminosity and machine conditions
that have made this work possible.
We acknowledge support from the
Natural Sciences and Engineering Research Council (Canada),
Institute of High Energy Physics (China),
Commissariat \`a l'Energie Atomique and
Institut National de Physique Nucl\'eaire et de Physique des Particules
(France),
Bundesministerium f\"ur Bildung und Forschung
(Germany),
Istituto Nazionale di Fisica Nucleare (Italy),
The Research Council of Norway,
Ministry of Science and Technology of the Russian Federation,
Particle Physics and Astronomy Research Council (United Kingdom), the
Department of Energy (US),
and the National Science Foundation (US). In addition, individual support 
has been received from the Swiss 
National Foundation, the A. P. Sloan Foundation, the Research Corporation,
and the Alexander von Humboldt Foundation.
The visiting groups wish to thank 
SLAC for the support and kind hospitality
extended to them.

%% file: paper.bbl
\begin{thebibliography}{99}

\bibitem{ref:sin2b}
\babar\ Collaboration,  B.\ Aubert {\em et al.},
``A study of time dependent asymmetries in $\Bz\rightarrow\jpsi\KS$ and 
$\Bz\rightarrow\psitwos\KS$ decays'', \babar-CONF-00/01,
submitted to the XXX$^{th}$ International Conference on High Energy
Physics, Osaka, Japan.

\bibitem{ref:babar}
\babar\ Collaboration, B.\ Aubert {\em et al.},
``The first year of the \babar\ experiment at \pep2'', 
\babar-CONF-00/17, submitted to the XXX$^{th}$ International
Conference on High Energy Physics, Osaka, Japan.

\bibitem{ref:incharm}
\babar\ Collaboration, B.\ Aubert {\em et al.},
``Inclusive $B$ decays to charmonium final states'', 
\babar-CONF-00/04, submitted to the XXX$^{th}$ International
Conference on High Energy Physics, Osaka, Japan.
.
\bibitem{ref:PDG}
Particle Data Group, D.\ E.\ Groom {\em et al.}, \epjc{15} (2000) 1.

\bibitem{ref:ARGUS}
ARGUS Collaboration, H.\ Albrecht {\em et al.}, \pl{B254} (1991) 288.

\bibitem{ref:physbook}
For a discussion of this technique, see
P.\ F.\ Harrison and H.\ R.\ Quinn, eds., ``The
\babar\ physics book'', SLAC-R-405 (1998) and references therein.

\bibitem{barlow}
R.\ Barlow, \nim{A297} (1990) 496.

\bibitem{CLEO97}
CLEO Collaboration, C.\ P.\ Jessop {\em et al.},
\prl{79} (1997) 4533.
\end{thebibliography}
